\documentclass[twocolumn,secnumarabic,amssymb,aps,prd,superscriptaddress]{revtex4}

\usepackage{graphicx,amssymb,amsmath}
\usepackage{epsfig}
\usepackage{hyperref}
\usepackage[usenames,dvipsnames]{color}
\usepackage{verbatim}

\newcommand{\nn}{\nonumber}
\newcommand{\eq}[1]{Eq.~\eqref{#1}}
\newcommand{\fig}[1]{Fig.~\ref{#1}}
\newcommand{\sctn}[1]{\S~\ref{#1}}
\newcommand{\appndx}[1]{Appendix~\ref{#1}}
\newcommand{\ie}{\textit{i.e.} }

\newcommand{\kbt}{k_\textmd{B}T}

\newcommand{\bsm}[1]{B_{\textmd{c},#1}}

\newcommand{\bhs}{\bsm{2}^\textmd{HS}}
\newcommand{\br}{\bsm{2}^*}
\newcommand{\rsm}{R_\textmd{c}}
\newcommand{\vsm}{V_\textmd{c}}

\newcommand{\rdep}{R_\textmd{d}}
\newcommand{\vdep}{V_\textmd{d}}
\newcommand{\phidep}{\phi_\textmd{d}}
\newcommand{\ndep}{N_\textmd{d}}
\newcommand{\udep}{u}
\newcommand{\uao}{\udep_\textmd{AO}}
\newcommand{\umtm}{\udep_\textmd{MT}}
\newcommand{\uaomin}{\udep_\textmd{AO}^\textmd{min}}

\newcommand{\udepcWCA}{\udep_\textmd{cWCA}}
\newcommand{\udepsWCA}{\udep_\textmd{sWCA}}
\newcommand{\uWCA}{U}
\newcommand{\uhs}{\uWCA_\textmd{HS}}
\newcommand{\ucWCA}{\uWCA_\textmd{cWCA}}
\newcommand{\usWCA}{\uWCA_\textmd{sWCA}}
\newcommand{\utot}{W}
\newcommand{\utotcWCA}{\utot_\textmd{cWCA}}
\newcommand{\utotsWCA}{\utot_\textmd{sWCA}}

\begin{document}

\title{Coarse-Grained Molecular Dynamics Simulations of Depletion-Induced Interactions for Soft Matter Systems}

 \author{Tyler N. Shendruk}
 \email{tyler.shendruk@physics.ox.ac.uk}
 \affiliation{The Rudolf Peierls Centre for Theoretical Physics, Department of Physics, Theoretical Physics, University of Oxford, 1 Keble Road, Oxford, OX1 3NP, United Kingdom}
 \author{Martin Bertrand}
 \author{James L. Harden}
 \author{Gary W. Slater}
 \affiliation{University of Ottawa, Department of Physics, 150 Louis-Pasteur, Ottawa, ON, K1N 6N5, Canada}
 \author{Hendrick W. de~Haan}
 \affiliation{University of Ontario Institute of Technology, Faculty of Science, 2000 Simcoe St. North, Oshawa, ON, Canada}
 \date{\today}

\begin{abstract}
Given the ubiquity of depletion effects in biological and other soft matter systems, it is desirable to have coarse-grained Molecular Dynamics simulation approaches appropriate for the study of complex systems. 
This paper examines the use of two common truncated Lennard-Jones (WCA) potentials to describe a pair of colloidal particles in a thermal bath of depletants. 
The shifted-WCA model is the steeper of the two repulsive potentials considered, while the combinatorial-WCA model is the softer. 
It is found that the depletion-induced well depth for the combinatorial-WCA model is significantly deeper than the shifted-WCA model because the resulting overlap of the colloids yields extra accessible volume for depletants.
For both shifted- and combinatorial-WCA simulations, the second virial coefficients and pair potentials between colloids are demonstrated to be well approximated by the Morphometric Thermodynamics (MT) model. 
This agreement suggests that the presence of depletants can be accurately modelled in MD simulations by implicitly including them through simple, analytical MT forms for depletion-induced interactions. 
Although both WCA potentials are found to be effective generic coarse-grained simulation approaches for studying depletion effects in complicated soft matter systems, combinatorial-WCA is the more efficient approach as depletion effects are enhanced at lower depletant densities. 
The findings indicate that for soft matter systems that are better modelled by potentials with some compressibility, 
predictions from hard-sphere systems could greatly underestimate the magnitude of depletion effects at a given depletant density. 
\end{abstract}

\maketitle

\section{Introduction}

Depletion forces of entropic origin are universal in soft matter systems, where they dictate the formation of a variety of soft solid phases~\cite{lekkerkerker,bibette99,tuinier03}. 
In particular, for biological systems, these forces have been shown to play a prominent role in a myriad of phenomena~\cite{marenduzzo04,marenduzzo06}. For example, we know depletion induced attractions promote fiber bundling~\cite{hosek04}, aggregation of red blood cells~\cite{steffen13}, and DNA collapse~\cite{pelletier12}. Reaction kinetics can also be influenced such as in DNA loop formation~\cite{marenduzzo06}, protein folding~\cite{zhou96,snir05}, and modified ligand-protein binding within nuclei~\cite{richter08}. Furthermore, these short-ranged depletion forces have been used to control biological systems, such as form active bundles~\cite{sanchez12}) and improved polymerase chain reactions~\cite{lareu07}.

Additionally, depletion forces occur in countless non-biological soft matter contexts. 
They are often used to engineer novel microscopic systems by inducing ordered structures~\cite{yodh01}, as is done to align tobacco mosaic viruses~\cite{adams98,glotzer07,li08}. 
Depletion forces can be significant enough to cause separation in colloid-polymer mixtures~\cite{tuinier03} and determine the phase diagrams of hard-sphere mixtures~\cite{roth00,roth10}. 
The presence of depletants can also alter the conformations of polymer chains~\cite{sear98,haya05}, form lock-and-key complexes~\cite{sacanna10,wang11,ashton13} and to control particle motion and modify curvature in vesicles~\cite{dinsmore96,dinsmore98}.

Yet, despite their experimental significance, simulations of complex situations that include depletion-induced interactions due to a thermal bath of smaller depletant particles are relatively rare. 
Computational research primarily focuses on the nature of the depletion forces themselves. 
A rich literature on binary hard-sphere fluids has benefited from Density Functional Theories (DFT) (see Refs.~\cite{schmidt03,roth10}). 
The majority of the work using DFT has been focused on understanding the statistical state of hard-sphere mixtures~\cite{roth99,roth00,schmidt00,roth01,oettel04,egorov04,roth06,konig06,herring07,oettel09}. 
Likewise, Monte Carlo (MC) simulations have played an essential role in verifying various depletion theories~\cite{biben96,dickman97,lue99,santos05}. 
The computational efficiency of MC simulations is continually improving~\cite{liu04,ashton13b} and they are being applied to more exotic mixtures than binary hard-spheres~\cite{li08,odriozola08,triplett10} and  material properties~\cite{rickman11}. 

Since such numerical calculations are computationally expensive, they have primarily been limited to simple situations. 
Furthermore, it is not always obvious how they can be used in conjunction with the generic particle-based computational tools that soft matter physicists have found to be useful for studying complex microscopic matter, such as Molecular Dynamics (MD)~\cite{slater09}. 
In order to study the physical phenomena that the depletants induce in complex soft matter systems, coarse-grained computational models that can seamlessly couple to those used by the soft matter community are needed. Previous work has used hard-core Yukawa potentials to study how additional interactions between depletants can impact depletion-induced effects~\cite{louis02}, while molecular simulations have been applied to better understand gelation in depletant-colloid systems~\cite{lue99} with large size ratios~\cite{henderson05} and the effects of using polymers as depletants~\cite{smith03,kim09,kim12}. 

\begin{figure}[tb]
 \begin{center}
  \includegraphics[width=0.4\textwidth]{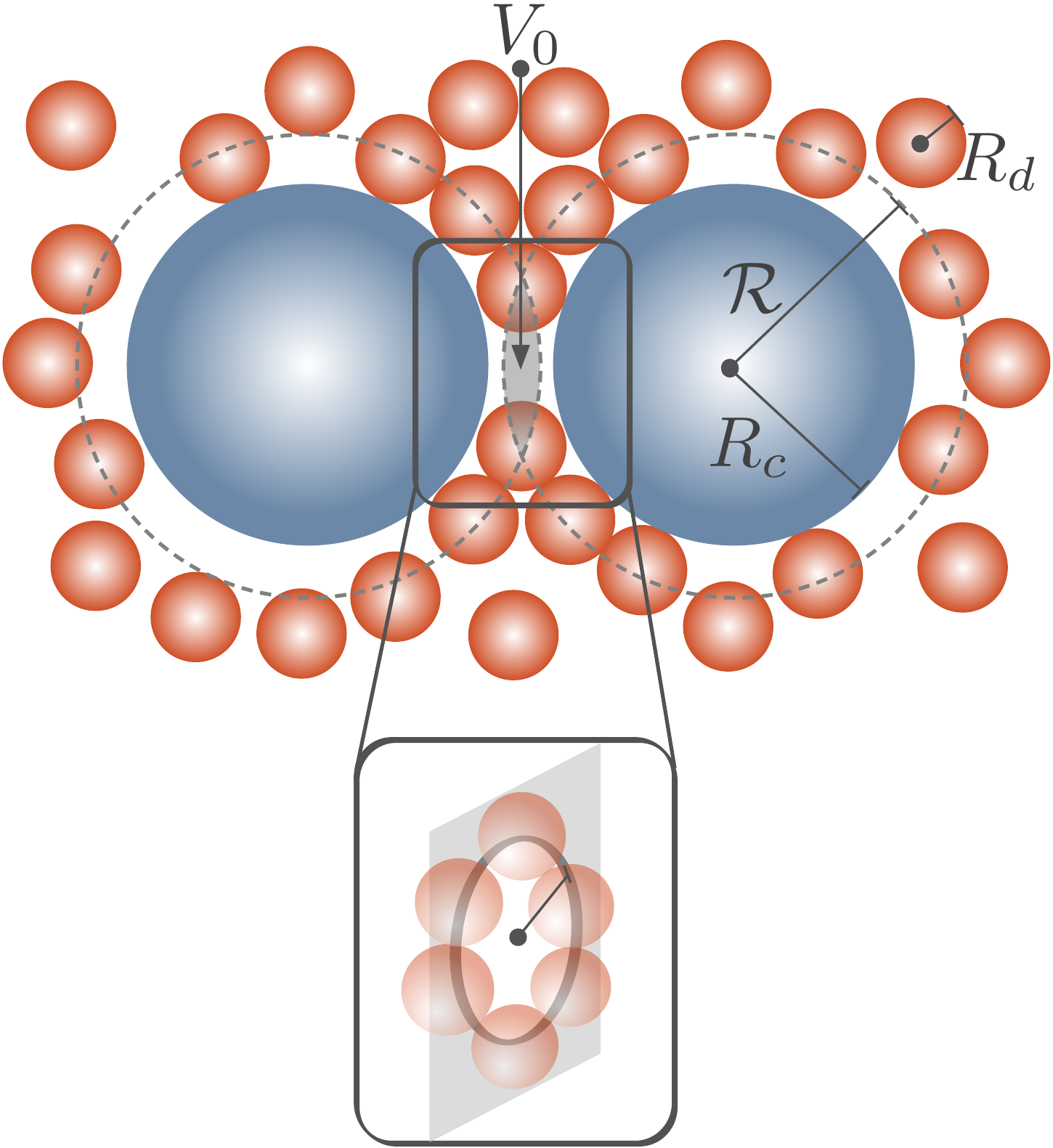}
  \caption{Two relatively large spherical colloidal particles (of radius $\rsm$) in a solution of depletants (radius $\rdep<\rsm$); each possesses an excluded volume of radius $\mathcal{R} = \rsm+\rdep$. 
  When the colloids are near ($2\rsm<r<2\mathcal{R}$) an overlap volume $V_o$ exists causing the number of states available to the depletants to be greater than when the colloids are separate ($r \geq 2\mathcal{R}$). 
  This leads to an effective force that favours bringing the colloids together. }
  \label{fig:overlap}
 \end{center}
\end{figure}

Kim and Szleifer~\cite{kim10} studied the attraction between two approaching polymer chains in a bath of depletants 
via MD simulations using a truncated Lennard-Jones potential, which is also called the Week-Chandler Andersen (WCA) potential.
The interaction was characterized by calculation of the potential of mean force (PMF),
but as the system was quite extensive in terms of the number of particles, only a small number of volume fractions of depletants could be studied.
In this work, we focus on an elementary system: two colloids in a bath of depletants (\fig{fig:overlap}). 
This allows us to explore depletant interactions across a wide range of volume fractions for two common WCA implementations.
In addition, the results are compared to theoretical predictions for the depletion potentials, including the Morphometric Thermodynamics model~\cite{oettel09,botan09,ashton11}.

Our work demonstrates that the details of the excluded volume interactions used in the MD simulations greatly impacts depletion-induced forces. 
Indeed, an appropriate choice of parameters can significantly increase the depth of the attractive well, 
which in turn decreases the second virial coefficient and effective volume of the colloids, in a predictable manner. 
As a result, simulations using ``soft'' yet steep potentials can reveal important entropic effects at lower volume fractions of depletants than hard sphere simulations and thus reduce computational costs for future studies into depletion effects in complex soft matter systems. 
Further, for soft matter systems in which the most appropriate coarse-grained model allows for some softness in the interaction between two constituents, 
the magnitude of depletion effects would be significantly underestimated by hard sphere models
and better approximated by potentials that are not infinitely steep at the nominal edge of the particle.

\subsection{Depletion-Induced Attraction}
Because they can produce ``order from disorder''~\cite{phillips09}, at first glance entropic forces can feel counter-intuitive. 
In fact, because they are purely entropic in nature, depletion interactions are universal in biology and common in soft matter systems. 
Before discussing the simulation techniques or results found in this work, let us first review depletion-induced pair potentials. 

As shown in \fig{fig:overlap}, consider two large, hard, colloidal spheres of radius $\rsm$ and volume $\vsm$ suspended in a thermal bath of small, hard depletant spheres of radius $\rdep$ and volume $\vdep$ with thermal energy $\kbt=\beta^{-1}$. 
The large colloids each have a thin sheath of radius $\mathcal{R} = \rsm+\rdep$ from which the centres of the depletants are excluded. 
When the two large spheres are brought together such that their centre-to-centre separation $r$ is less than $2\mathcal{R}$, their excluded volumes overlap and the free space available to the depletants increases. 
This results in a free energy change as a function of available volume for the depletants giving rise to an osmotic pressure pushing the colloids together. 

\subsection{Asakura-Oosawa Pair Potential}
Asakura and Oosawa were the first to estimate this depletion-induced pair potential~\cite{asakura54}. 
They discussed the attraction in terms of an entropic, depletant-induced (and therefore statistical in nature), short-ranged pair potential $\udep \left(\rho_{dep}, r\right)$. 
The Asakura-Oosawa (AO) model has also been referred to as the penetrating hard spheres (PHS) model~\cite{lekkerkerker} because the depletants are assumed to be hard with respect to the large colloids but an ideal gas amongst themselves, which is only appropriate for small volume fractions of depletants. 
In the thermodynamic, dilute limit of depletants ($\ndep \rightarrow \infty$ while $\phidep \ll 1$), one can approximate the osmotic pressure to be given by a van't Hoff relation $ \Pi \approx -\left(\phidep/\vdep\right) \beta^{-1} $. 
Explicitly, the AO entropic depletion pair potential is 
\begin{align}
 \beta \uao &=
  \begin{cases}
   -\left(\frac{ \mathcal{R}}{\rdep}\right)^3 \left[ 1 - \frac{3}{4}\frac{r}{\mathcal{R}} + \frac{1}{16}\left(\frac{r}{\mathcal{R}}\right)^3\right] \phidep & 2\rsm \leq r \leq 2\mathcal{R}\\
   0 & r>2\mathcal{R}
  \end{cases}
 \label{aoPot}
\end{align}

\begin{figure}[tb]
 \begin{center}
  \includegraphics[width=0.5\textwidth]{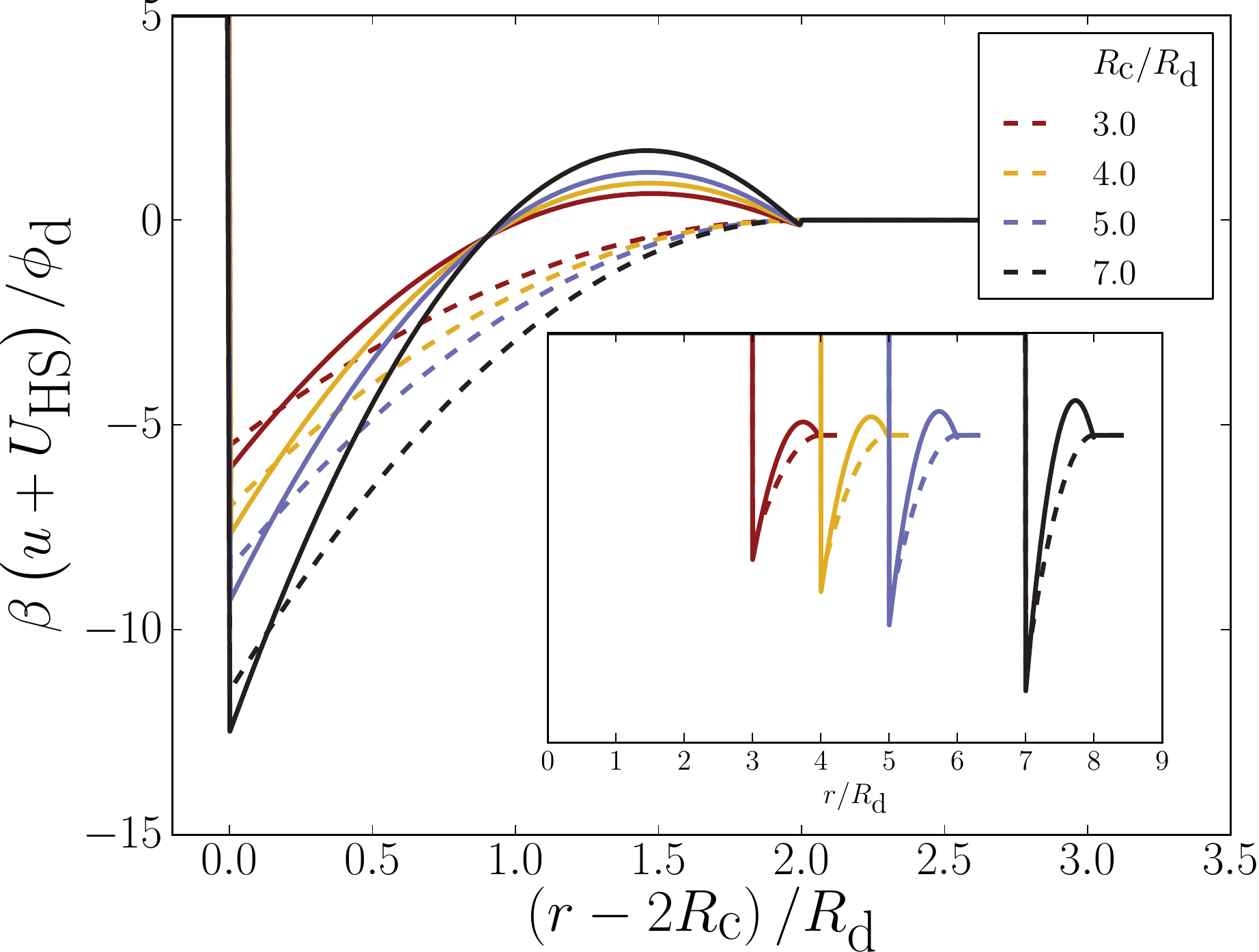}
  \caption{Depletion-induced attractive pair potential between two spherical, hard colloids for a variety of colloid-depletant ratios. 
  The inset shows the entropic pair potentials for various hard-sphere colloid and depletant sizes, while the main figure shifts these curves. 
  The main abscissa is a shifted and scaled by the centre-to-centre separation emphasizing differnces in well depths. 
  The hard-sphere Aakura-Oosawa (A0) model (\eq{aoPot} and \eq{hs}) is shown as dashed lines, while the Morphometric Thermodynamics (MT) predictions are shown as solid lines.
  Scaling the pair potentials removes $\phidep$ dependence from the AO model but not the MT model. 
  The MT curves shown are for the example volume fraction of $\phidep=0.2$ 
  }
  \label{fig:interactions}
 \end{center}
\end{figure}

Note that the AO model (and MT model, to be discussed below) is generally defined for hard spheres and must be supplemented with a steric-repulsion condition for $r<2\rsm$. 
For hard spheres of radius $\rsm$,
\begin{small}
\begin{align}
 \uhs &=
 \begin{cases}
  \infty  & r <  2\rsm\\
   0 & r \geq 2\rsm.
  \end{cases}
  \label{hs}
\end{align}
\end{small}
Hence, the total interaction is $\uao+\uhs$ in this definition.
As will be shown, we are free to relax this constraint when the particles are not strictly hard spheres.

The AO pair potential (\fig{fig:interactions}; dashed lines) is  characterized by four points: 
(i) the potential is strictly attractive; 
(ii) the range is set by the depletant size $\rdep$; 
(iii) the minimum occurs when the large colloids are in contact 
\begin{align}
 \beta\uaomin &= - \left[1+\frac{3}{2}\frac{\rsm}{\rdep}\right] \phidep,
  \label{min}
\end{align}
which deepens as the size ratio $\rsm/\rdep$ is increased (though is not zero in the limit $\rsm/\rdep\rightarrow1$~\cite{zaccone12}); 
(iv) via osmotic pressure, the volume fraction of depletants and temperature both scale the well depth.

The AO model is beautifully simple; however, it has two major limitations:
\begin{itemize}
 \item Depletants must be small $\rdep\ll\rsm$. 
 \item The volume fraction must be small $\phidep \ll 1$.
\end{itemize}
The essential conclusion is that in order to consider the many interesting situations that do not meet these conditions, we must move beyond the AO limit.

\subsection{Morphometric Pair Potential} \label{mtm}
To move beyond the $\phidep \ll 1$ limit, the free energy cost of inserting the large colloids must be greater than it is in the PHS assumption. 
Simply replacing the PHS osmotic pressure with a form that includes higher virial coefficients for the hard sphere gas of depletants allows the AO model to predict colloid pressures at higher $\phidep$ but is not sufficient to predict the form of the depletion pair potential~\cite{lekkerkerker90,gotzelmann98}. 
The missing ingredient is the realization that colloids that are close to one another mutually perturb the statistical arrangement of their surrounding shells of depletants and cause an inhomogeneous probability distribution of depletants. 
This results in the pair potential oscillating between attractive and repulsive as a function of separation~\cite{gotzelmann98}. 
The well effectively becomes deeper as it is now the difference between the maximum due to repulsion and the minimum due to attraction, but the minimum itself remains mostly unaffected. 

Computationally intensive numerical techniques such as DFT can determine the local depletant particle density and so find the resulting entropic pair potential between colloids. 
Such techniques have proven extremely successful in exploring non-additive effects~\cite{roth01} but are not straight-forward analytically simple models. 
One conceptually simple model that can reproduce the first repulsive/anti-correlation component of the pair potential is called Morphometric Thermodynamics (MT)~\cite{oettel09,botan09,ashton11}.

The MT model discusses free energy in terms of the geometric shapes involved and their thermodynamic conjugates. 
In the AO model, only the overlap volume $V_o$ and osmotic pressure $\Pi$ associated with it are considered. 
Just as the osmotic pressure is a thermodynamic force that arises because of restrictions on the placement of depletants within the overlap volume, an entropic surface tension arises due to the cost of disallowing depletants to be at certain points on the shell of depletants surrounding each colloid due to the presence of the other colloid (\fig{fig:overlap}). 
The entropic interaction pair potential $\udep$ is then modelled as arising from changes to the accessible volume $V_o$, to the surface area $A$ restriction with resulting entropic surface tension $\gamma$ and to the Gaussian curvatures of the overlap surface $C_1$ and $C_2$ with corresponding entropic bending rigidities $\kappa_1$ and $\kappa_2$:
\begin{align}
 \umtm &= \Pi V_o + \gamma A + \kappa_1 C_{1} + \kappa_2 C_{2}
 \label{morpho}
\end{align}
where the geometric terms are functions of the colloid-colloid separation while the prefactors depend on the volume fraction of depletants (listed in Appendix B). 
As shown in Fig. \ref{fig:interactions} (solid lines), the MT framework is able to produce the attractive well and also the primary repulsive barrier, while the AO model only predicts the attractive well. 
Correlations and anti-correlations beyond the primary contributions cannot be reproduced within the MT model. 

\section{Simulation Methods}\label{sims}

Simplified analytical models such as the AO or MT potentials are appealing for straightforward systems but coarse-grained simulations offer a path to studying more complicated soft matter systems. 
In order to computationally study depletion effects via Molecular Dynamics simulations, a generic steep repulsive potential to produce hard steric effects between colloids and depletants is needed. 
To coarse-grain the system, we neglect all hydrodynamic interactions that may propagate through the fluid medium~\cite{balfaluy93,padding06} and any possible non-equilibrium effects~\cite{dzubiella03}. 
We consider the solvent to be simply a dissipative thermal bath with Langevin Dynamics. 
Both the colloidal particles and the depletants are modelled as quasi-``hard spheres'' in the framework of coarse-grained MD. 

Because depletion forces arise due to excluded volumes between particles that are inaccessible to the smaller depletant particles, any repulsive potential can technically be used to represent the steric repulsion (though as we shall see, the details can have a significant impact on the magnitude of the effective interactions). 
In typical coarse-grained MD simulations, this is done using the Lennard-Jones (LJ) pairwise potential, which includes two terms, a steric repulsion that goes as $r^{-12}$ and a dispersion-type interaction that goes as $r^{-6}$~\cite{slater09}. 
In order to neglect attractive dispersion-type attraction, the LJ potential is commonly truncated at its minimum and shifted to positive values, as will be explained in \sctn{sctn:sWCA} and \sctn{sctn:cWCA}. 
This truncated LJ potential is referred to as the Weeks-Chandler-Andersen (WCA) potential. 
This is a simple and commonly employed potential; however, using a potential that is not infinitely steep alters the attractive entropic interactions and they must be carefully characterized. 
It will be seen in this work that an appropriate choice of parameters can significantly increase the depth of the depletion induced potential well, which in turn decreases the second virial coefficient and the effective volume of the colloids. 
We consider two versions of the WCA potential as coarse-grained generic simulation models for depletion effects.
For both approaches, the characteristic energy for all interactions is $\epsilon$ and the characteristic length scale is $\sigma$ --- which is equal to the diameter of the depletants.

\subsection{Shifted WCA} 
\label{sctn:sWCA}
\begin{figure}[tb]
 \begin{center}
  \includegraphics[width=0.45\textwidth]{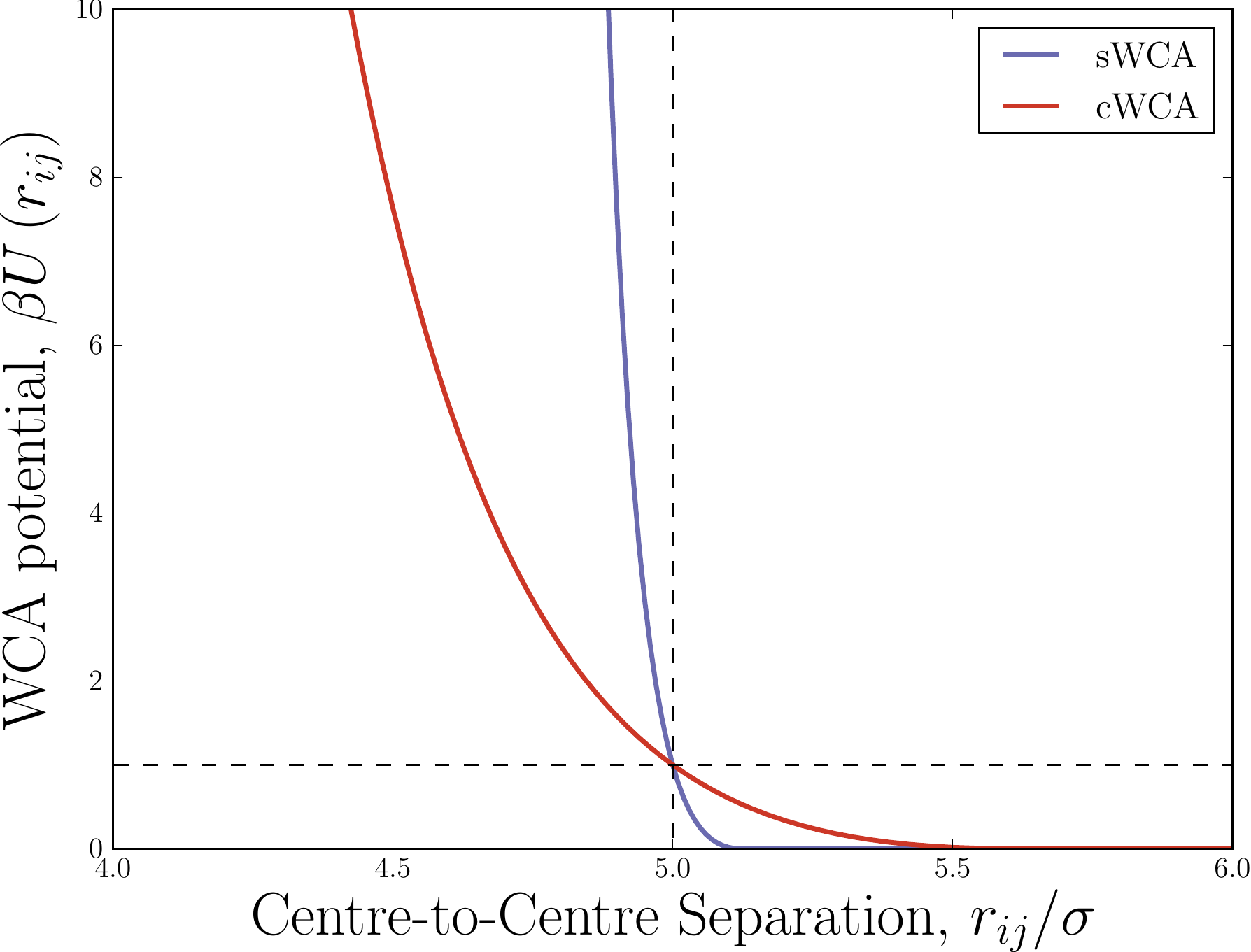}
  \caption{Comparison of the shifted-WCA (blue) and combinatorial-WCA (red) potentials for $\rsm=2.5\sigma$, $\varepsilon=1\kbt$ and $\Delta_{ij}=4\sigma$ for the shifted-WCA case.   
  The combinatorial-WCA extends further and diverges at a small separation and is thus the ``softer" of the two.
  However, both potentials are hard in that they rise as $1/r^{12}$.} 
  \label{fig:pot_plot}
 \end{center}
\end{figure}

The shifted-WCA model (sWCA) associates the same repulsive potential to each spherical particle but shifts the barrier depending on their radii. 
For a particle of nominal radius $R_i$ and one of $R_j$, it is given by
\begin{align}
 \uWCA &=
 \begin{cases}
   \infty & r_{ij} \leq \Delta_{ij} \\ 
   4\varepsilon\left[ \left(\frac{\sigma}{r_{ij} -\Delta_{ij}}\right)^{12} - \left(\frac{\sigma}{r_{ij}-\Delta_{ij}}\right)^{6}\right]  + \varepsilon& \Delta_{ij} < r_{ij} < r^\textmd{cut}_{ij} \\
   0 & r_{ij} \geq r^\textmd{cut}_{ij},
  \end{cases}
 \label{eq:sWCA}
\end{align}
where $r_{ij}$ is the centre-to-centre separation between particles $i$ and $j$, 
$\varepsilon$ is the depth of the potential well, 
$\sigma$ is the nominal lengthscale of the potential, 
$\Delta_{ij} = R_{i} + R_{j} - \sigma$ is a shift of the potential and 
$r^\textmd{cut}_{ij} = 2^{1/6}\sigma + \Delta_{ij}$ is the cutoff distance. 
The repulsive potential thus rises in the same rapid manner, over the same distance of $2^{2/6}\sigma$, for all pairs regardless of the size of the particles. 
Here we have chosen $\sigma$ and $\varepsilon$ in \eq{eq:sWCA} to explicitly be the characteristic values but one could conceivably chose some other distance and/or energy.
As shown in Fig. \ref{fig:pot_plot}, the shifted-WCA potential $\usWCA$ rises from zero at $r^{\textmd{cut}}_{ij}$, crosses $\usWCA=1\varepsilon$ at $r_{ij}=R_{i} + R_{j}$, and diverges as $r \rightarrow \Delta_{ij}$.

\subsection{Combinatorial WCA}
\label{sctn:cWCA}
On the other hand, the combinatorial-WCA model (cWCA) does not shift the WCA repulsion but rather sets an effective size between two MD particles. 
It is defined by
\begin{align}
 \uWCA &=
 \begin{cases}
   4\varepsilon \left[ \left(\frac{\sigma_{ij}}{r_{ij}}\right)^{12} - \left(\frac{\sigma_{ij}}{r_{ij}}\right)^{6}\right]  + \varepsilon&  r_{ij} < r^\textmd{cut}_{ij} \\
   0 & r_{ij} \geq r^\textmd{cut}_{ij}.
  \end{cases}
 \label{eq:cWCA}
\end{align}
where $\sigma_{ij}$ is no longer constant but instead is given by $R_{i} + R_{j}$ and $r^\textmd{cut}_{ij} = 2^{1/6}\sigma$. 
The resulting repulsion between particles $\ucWCA\left( r_{ij} \right)$ is still hard --- the potential still steeply rises as $\sim r_{ij}^{-12}$.  
However, the combinatorial-WCA repulsion between two colloids is slightly softer than its shifted-WCA counterpart. 
As shown in \fig{fig:pot_plot}, the repulsive potential extends further and does not rise as sharply. 
The same is true for colloid-depletant interactions.
Since $\sigma$ is equal to the size of the depletants, $2\rdep$, the depletant-depletant interactions are equivalent in the sWCA and cWCA potentials. 

\subsection{Effective Colloid Size}
In using continuous potentials instead of perfectly hard spheres, the question arises as to what the effective size of the colloids are.
Not only are the potentials not infinite at the nominal radius, but the colloid-colloid interaction extends beyond the $2\rsm$ (\fig{fig:pot_plot}).
The second virial coefficient for a pairwise-interacting solution of colloids can be used to define an effective volume (and hence radius) for the colloids defined by either sWCA and cWCA: 
\begin{align}
 \bsm{2} &= -2\pi\int_0^\infty \left( e^{-\beta \uWCA\left(r\right)} - 1 \right) r^2 dr.
 \label{method_2nd}
\end{align}
Here $\uWCA\left(r\right)$ is the potential that defines the colloids, which for hard spheres is given by \eq{hs}. 
Hence for hard-spheres, the second virial coefficient is 4 times the volume of the hard sphere $\bhs = 4 \left( \tfrac{4}{3} \pi R_C^3 \right) = 4 V_{HS}$. 

For the sWCA potential, we use \eq{eq:sWCA} for colloids of nominal size $2\rsm$ in \eq{method_2nd} to solve for $\bsm{2}$.
Defining the effective volume of the sWCA colloids as $V_{sWCA}=\bsm{2}/4$, we can calculate $V_{sWCA}/V_{HS}$, which depends on the nominal radius $\rsm$. 
For the case of $\rsm=2.5 \sigma$ as studied in this work, we find that
\begin{align}
\frac{V_{sWCA}}{V_{HS}} = 1.010.
\end{align}
Hence, using the sWCA potential to define a colloid with a nominal diameter of $5\sigma$ results in a colloid that has a $1\%$ larger volume than the hard sphere equivalent.
This corresponds to an effective radius of $R_{sWCA}=2.508$. 

Following a similar procedure for the cWCA potential but using \eq{eq:cWCA}, we obtain
\begin{align}
\frac{V_{sWCA}}{V_{HS}} = 1.053.
\end{align}
The cWCA potential thus results in a colloid that has a 5.3\% larger volume than the hard sphere result, corresponding to an effective radius of $R_{cWCA}=2.543$.
As opposed to the sWCA result, the length of the cWCA potential scales with the nominal size and thus this result is independent of $\rsm$.

The $R_{sWCA}$ and $R_{cWCA}$ values given above are used in theoretical calculations in the rest of this work.
Although the corrections to the nominal radii appear to be small, they do have a significant effect on the results
and yield much better agreement between the simulations and the theory for the pair potentials, well depths, and $\bsm{2}$ values.
It is important to keep in mind that both of the WCA potentials are ``hard'' in that they are both short-ranged, rapidly rising, repulsive potentials. 

\subsection{Net potential between two colloids} 
The net pair potential between two colloids is simply the sum of the excluded volume potentials $U$ (\eq{eq:sWCA} or \eqref{eq:cWCA}) and the entropic component $u$:
\begin{align}
 \utot &= \uWCA\left(r_{ij},\rsm,\sigma\right) + \udep\left(\phidep,r_{ij},\rsm,\rdep\right). 
 \label{eq:net}
\end{align}
In this notation, $W$ indicates the net potential, $U$ denotes the repulsive, excluded volume potentials (WCA), and $u$ specifies the entropic, depletion potentials. 
When the WCA potentials are subtracted off the net interaction potential then only the depletion-induced component remains. 

\section{Simulations of Depletant-Induced Pair-Potential} \label{pairSims}

To measure the entropic pair potential between two colloids, the mean force on each colloidal particle held at a fixed separation is measured. 
Integrating over the entire range of simulated separations produces the potential of mean force $W^{\textmd{sim}}$ between two colloidal particles.

\subsection{Shifted-WCA}\label{sWCA:sims}

\begin{figure}[tb]
 \begin{center}
  \includegraphics[width=0.5\textwidth]{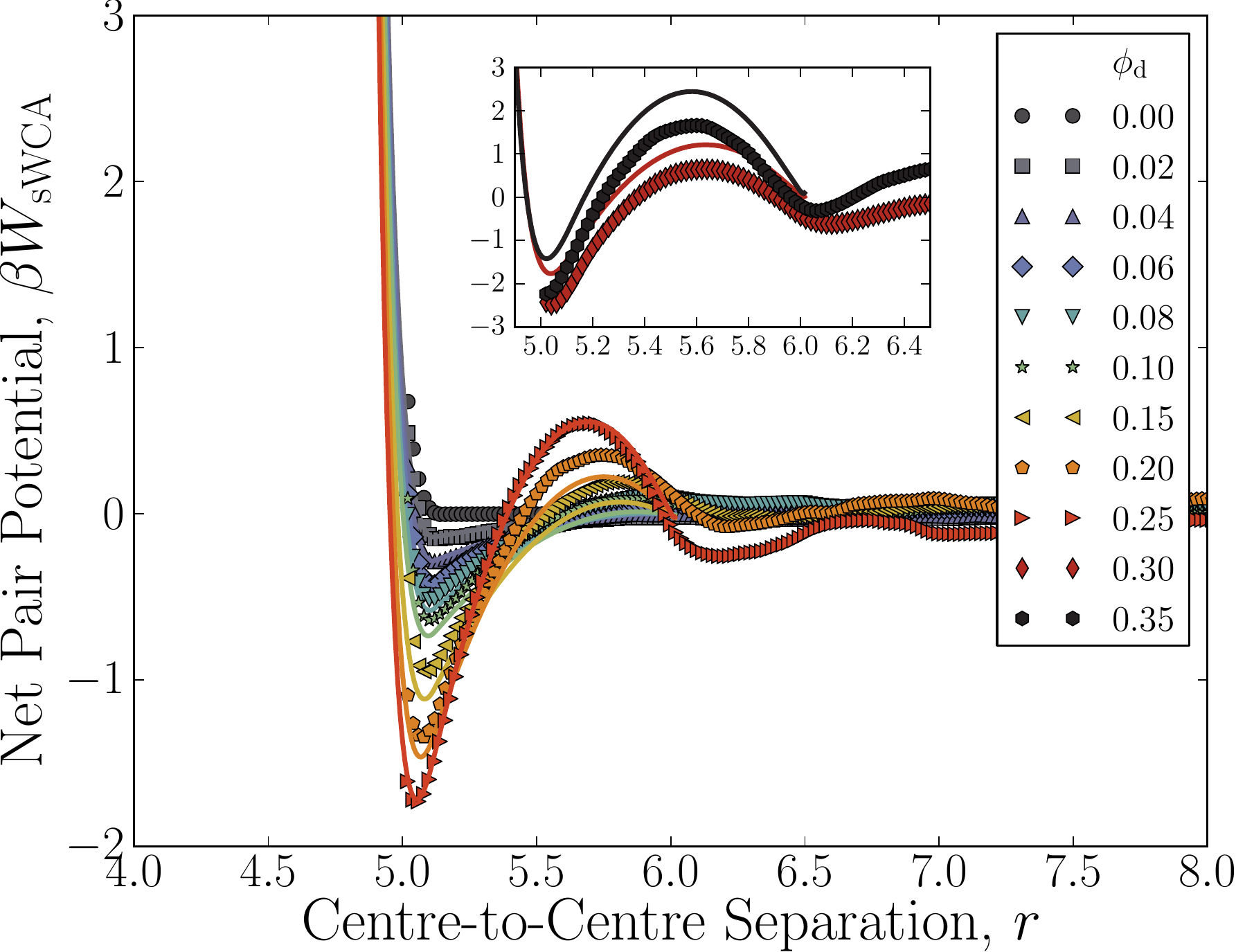}
  \caption{Simulated net pair potentials $W_\textmd{sWCA}^{\textmd{sim}}$ from integrated force measurements of shifted-WCA beads of size $\rsm=2.5\sigma$ in baths of various volume fractions of shifted-WCA depletants of size $\rdep=0.5\sigma$. 
  While the main figure shows low to moderate values, the inset shows the high volume fraction cases. 
  Solid lines are the sum of the entropic pair potential predicted by the Morphometric Thermodynamics (MT) model and the shifted-WCA repulsion, $u_\textmd{MT} + U_\textmd{sWCA}$. 
  }
  \label{fig:tot_sWCA}
 \end{center}
\end{figure}

\begin{figure}[tb]
 \begin{center}
  \includegraphics[width=0.5\textwidth]{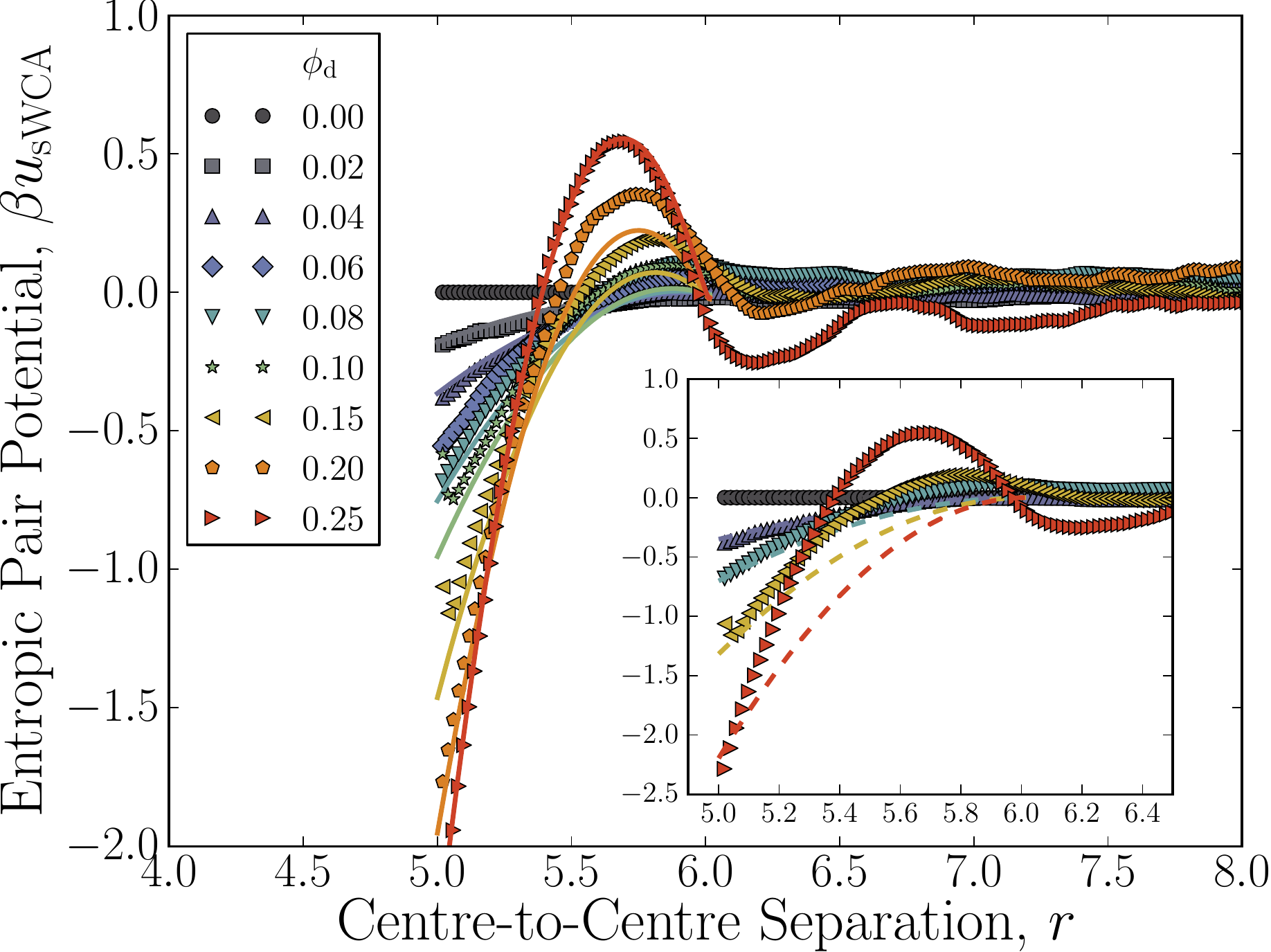}
  \caption{Simulated depletant-induced entropic pair potentials $\udepsWCA$ from integrated force measurements of shifted-WCA beads of size $\rsm/\rdep=5$. 
  Solid lines show the induced pair potentials predicted by the Morphometric Thermodynamics (MT) model, while dashed lines in the inset show the Aakura-Oosawa (AO) model. 
  These theoretical curves are zero for $r>2\mathcal{R}=6\sigma$ and infinite for $r<2\rsm=5\sigma$.
  }
  \label{fig:dep_sWCA}
 \end{center}
\end{figure}

The net pair potential $W_\textmd{sWCA}^{\textmd{sim}}\left(r\right)$, as measured by force integration, is shown in \fig{fig:tot_sWCA}. 
The strong, short-ranged, shifted-WCA repulsion $\usWCA$ is clearly visible at small separations and the potential diverges at $r=\Delta_{ij}=4\sigma$ for these simulations. 
For clarity, The main figure shows low to moderately volume fractions ($\phidep \leq 0.25$) and the inset shows the highest volume fractions considered here ($\phidep = 0.3$ and $0.35$). 
Subtracting $\usWCA$ off leaves only the depletion-induced component $\udepsWCA$ (\fig{fig:dep_sWCA}). 
Persistently negative potentials for large centre-to-centre separations suggests that the force integrated results from simulations underpredict the net pair potentials at the highest volume fractions of depletants considered ($\phidep>0.25$). 

At low volume fractions, the simulated pair potentials are well approximated by both the AO and the MT models but as $\phidep\gtrsim0.1$ the AO model begins to deviate from the measured curves (\fig{fig:dep_sWCA}; inset). 
On the other hand, the MT model continues to capture the behaviour of the primary attractive well and repulsive barrier (the region within approximately two depletants from the colloid's surface: $2\rsm \leq r \lesssim 2\left[\rsm + 2\rdep\right]$). 
Of course, the MT model does not capture the secondary features in \fig{fig:dep_sWCA} because current functionals are not applicable beyond $2\mathcal{R}=2\left(\rsm+\rdep\right)$~\cite{botan09}. 

\begin{figure}[tb]
 \begin{center}
  \includegraphics[width=0.5\textwidth]{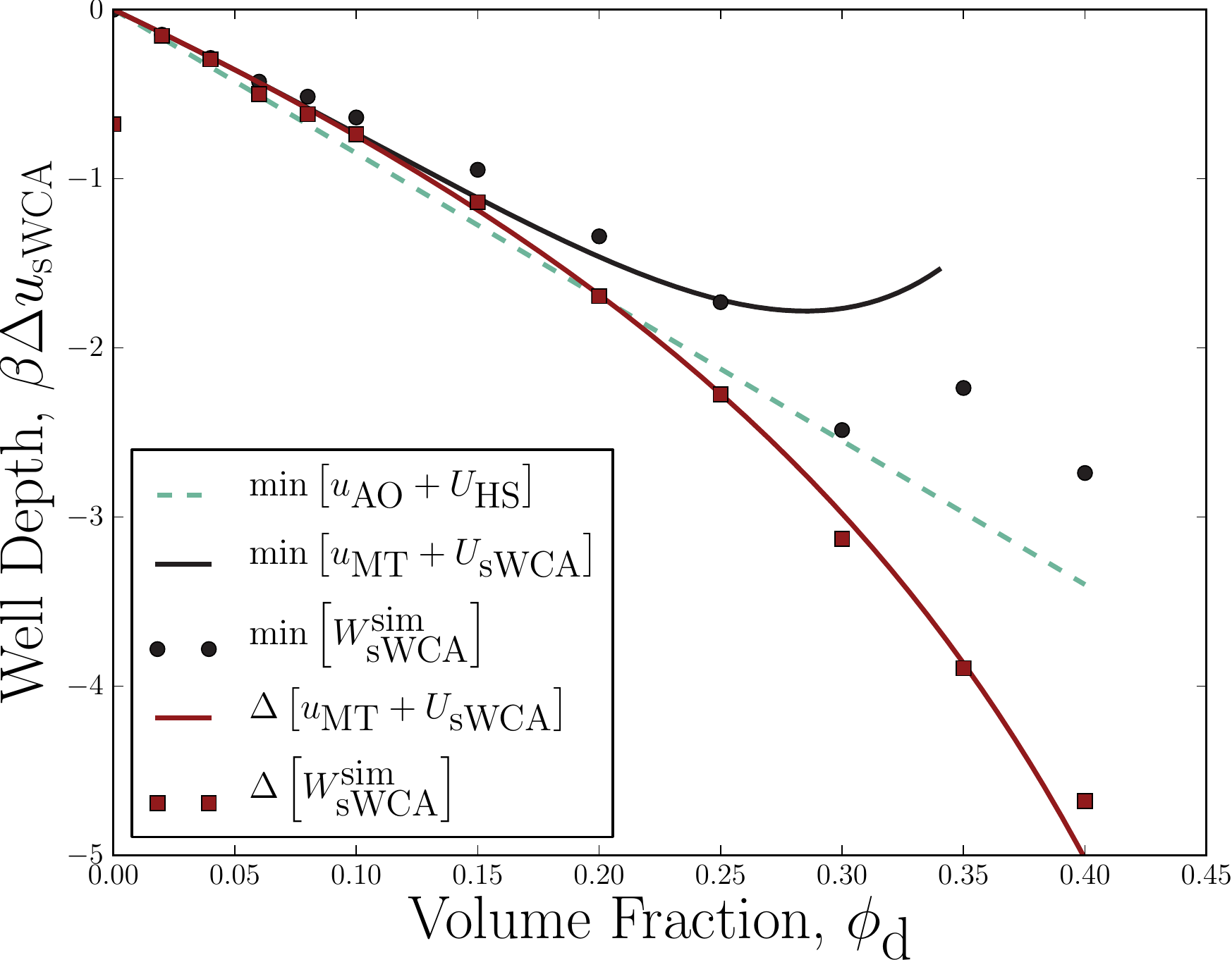}
  \caption{
  Measures of the well depth of the pair potential as a function of depletant volume fraction for shifted-WCA particles of size $\rsm=2.5\sigma$ and depletants of size $\rdep=0.5\sigma$. 
  The minimum from simulations ($\utotsWCA$; black circles) is compared to the Aakura-Oosawa (A0) value ($\uaomin$ from \eq{min}; dashed line) and the theoretical minima of the Morphometric Thermodynamics (MT) model plus the shifted-WCA potential (min[$u_\textmd{MT}+U_\textmd{sWCA}$]; solid black line). 
  The effective well depths (primary minimum minus primary maximum) from simulations ($\Delta\utotsWCA$; red squares) 
  and the theoretical net value ($\Delta[u_\textmd{MT} + U_\textmd{sWCA}$]; solid red line) are also shown. 
  }
  \label{fig:minHard}
 \end{center}
\end{figure}

Although, the AO model does not well represent the pair potential as a function of separation, \fig{fig:dep_sWCA} shows that it does estimate the contact energy well, \ie $u_\textmd{sWCA}^\textmd{min} \approx \uaomin$. 
This can be seen quantitatively in \fig{fig:minHard}. 
In fact, \fig{fig:minHard} shows that the linear AO minimum (\eq{min}) agrees with the simulated min[$W_\textmd{sWCA}^\textmd{sim}$]. 

To better understand these predictions, we theoretically combine the MT model for the depletion potential and the sWCA excluded volume potential: $u_\textmd{MT} + U_\textmd{sWCA}$.
This approach successfully replicates the net pair potential as compared to simulation for $\phidep\lesssim0.25$ (\fig{fig:tot_sWCA}; main). 
This is because the depletion potential is well approximated by the MT model for all but the highest depletant volume fractions (\fig{fig:dep_sWCA}). 
For the highest volume fractions considered (\fig{fig:tot_sWCA}; inset), the MT model appears to overpredict the primary barrier height and underpredict the primary well depth for sWCA. 
The minima of the $u_\textmd{MT} + U_\textmd{sWCA}$ data are shown in Fig. \ref{fig:minHard} as solid lines. 
While good agreement is found up to $\phi = 0.25$, min[$u_\textmd{MT} + U_\textmd{sWCA}$] actually begins to curve upwards thereafter.
This trend is an artifact of the MT model where at high volume fractions, the first repulsive peak is exaggerated --- which drags the curve upwards at $r \approx 2R_c$ and thus increases the minimum~\cite{ashton11}. 
Due to this artifact, even the linear AO model does a better job than the MT model at predicting the minimum in this regime. 

To correct this, the difference between the minimum of the contact attraction and the maximum of the anti-correlation barrier can be calculated.
This calculation can be done for both the simulations (\fig{fig:minHard}; red squares) and the $u_\textmd{MT} + U_\textmd{sWCA}$ potential
(this is not possible for the AO potential as no repulsive barrier is predicted).
Excellent agreement is now found between the simulation results and the theoretical $u_\textmd{MT} + U_\textmd{sWCA}$ potential across all studied volume fractions. 
This suggests that MT's overprediction of the primary barrier and underprediction of the primary well depth cancel. 

The well depth competes with thermal energy in determining if colloids prefer to agglomorate in clusters or explore the entire available volume. 
Qualitatively speaking since the well depth $\utotsWCA^\textmd{min}$ in \fig{fig:minHard} is only a few $\kbt$ even as $\phidep\rightarrow0.4$, we expect that extremely high number densities of depletants would be required to overcome thermal jostling. 
This will be further quantified when we consider the second virial coefficients of colloids in thermal baths of depletants (\sctn{2ndvirial}). 

\subsection{Combinatorial-WCA}\label{cWCA:sims}

\begin{figure}[tb]
 \begin{center}
  \includegraphics[width=0.5\textwidth]{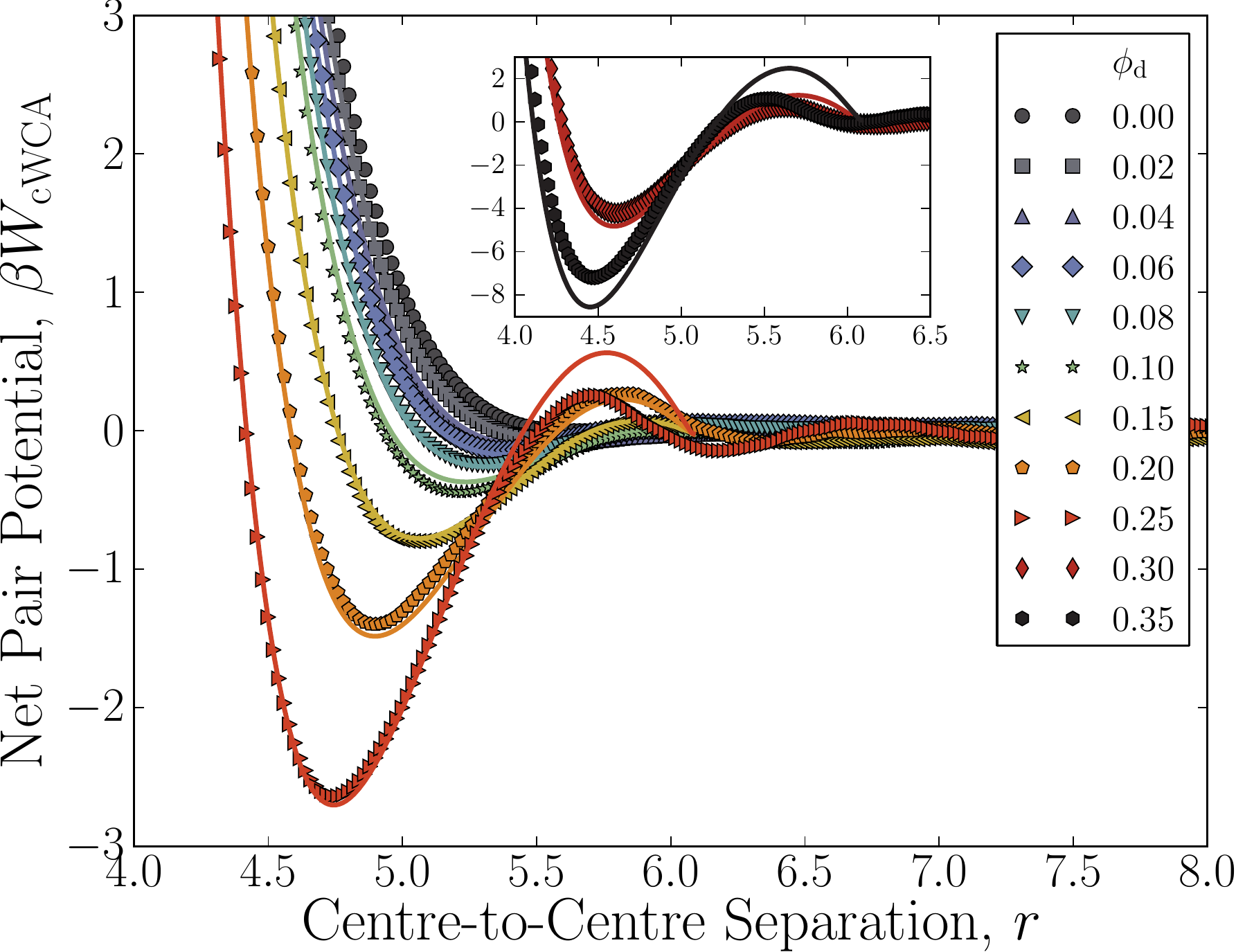}
  \caption{
  Simulated net pair potentials $\utotcWCA$ from integrated force measurements of combinatorial-WCA beads of size $\rsm=2.5\sigma$ in baths of various volume fractions of combinatorial-WCA depletants of size $\rdep=0.5\sigma$. 
  The inset shows the high volume fraction cases and the main figure shows the low to moderate values. 
  Solid lines are the sum of the MT model for the depletion potential and the combinatorial-WCA excluded volume potential, $u_\textmd{MT} + U_\textmd{cWCA}$.
  }
  \label{fig:tot_cWCA}
 \end{center}
\end{figure}

\begin{figure}[tb]
 \begin{center}
  \includegraphics[width=0.5\textwidth]{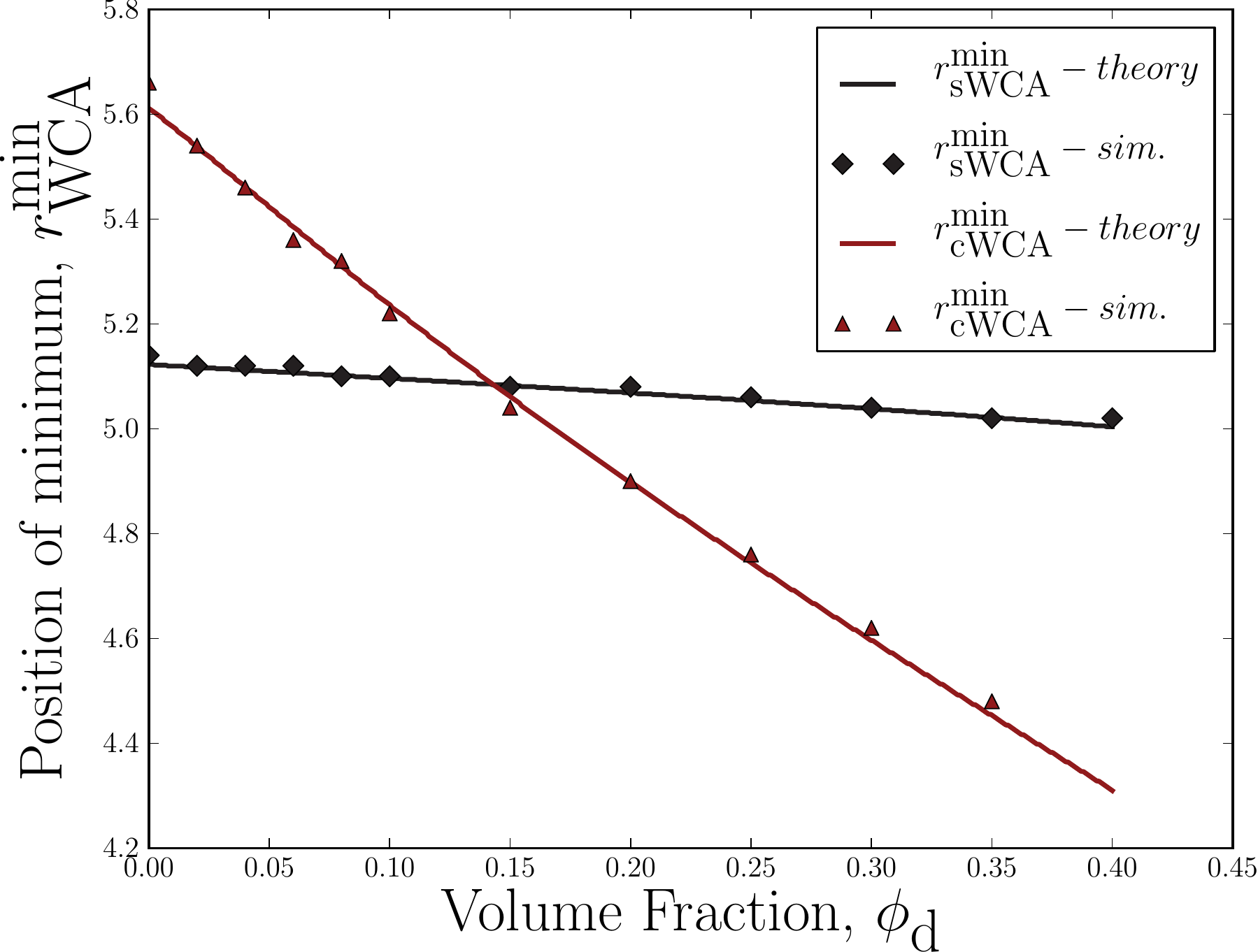}
  \caption{Position of the minimum in the net potential measured from simulations for both sWCA (black diamonds) and cWCA (red triangles).
  The prediction from the theoretical potentials $u_{MT} + U_{sWCA}$ (solid black line) and $u_{MT} + U_{cWCA}$ (solid red line) are also included.
  }
  \label{fig:min_pos}
 \end{center}
\end{figure}

The depletant-induced pair potential is measured in the same manner as in \sctn{sWCA:sims} for cWCA interactions; however, separations smaller than $2\rsm$ are allowed by the cWCA repulsion and must now be considered as seen in \fig{fig:tot_cWCA}. 
The competition between $\ucWCA$ and the entropic pair potential $\udepcWCA$ results in a minimum in the net interaction $\utotcWCA$ for all volume fractions. 
A comparison between \fig{fig:tot_sWCA} and \fig{fig:tot_cWCA} shows that the well is much deeper in the combinatorial-WCA model than for the shifted-WCA. 
This is because of the additional overlap that can occur. 
Since the $\ucWCA$ repulsion rises less dramatically and does not have an infinitely-hard repulsive core, the combinatorial-WCA wells are broader, extending to smaller separations.
For high volume fractions, the combinatorial-WCA repulsion $\ucWCA$ does not dominate over the depletion-induced pair potential $\udepcWCA$ until much smaller separations than in the shifted-WCA model. 
This can be seen in \fig{fig:min_pos}, which displays the location of the minimum of the net potentials $W_\textmd{cWCA}$ and $W_\textmd{sWCA}$ as measured in simulations as a function of volume fraction.
Excellent agreement is again found between theory (MT) and simulation.

At low volume fraction, colloids are actually more likely to be found further apart for cWCA than sWCA since the cut-off for the former extends further.
However, due to the relative softness of $u_\textmd{cWCA}$, the minimum in the potential decreases at a much faster rate and, at high volume fractions, 
colloids defined by cWCA are most likely found to be significantly closer together than for sWCA.
It is interesting to note that the dependence of $r_\textmd{min}$ on $\phidep$ is approximately linear for both WCA potentials,
with a cross-over around $\phidep \approx 0.14$ and $r_\textmd{min} \approx 5.1$. 

\begin{figure}[tb]
 \begin{center}
  \includegraphics[width=0.5\textwidth]{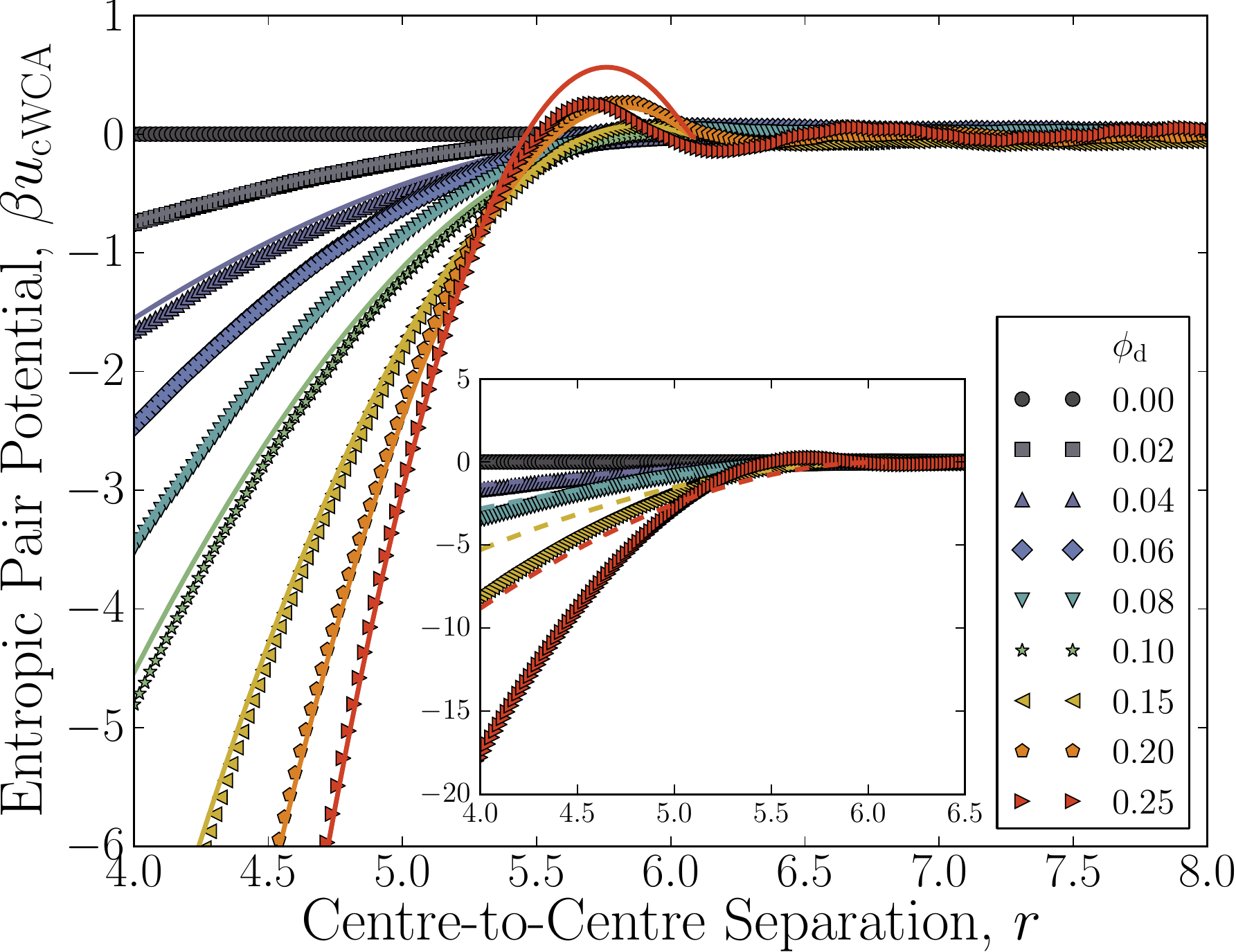}
  \caption{Simulated depletant-induced entropic pair potentials $\udepcWCA$ as a function of centre-to-centre separation for $\rsm/\rdep=5$. 
  Solid lines show the depletion-induced pair potentials predicted by the Morphometric Thermodynamics (MT) model, while dashed lines in the inset show the Aakura-Oosawa (AO) model. 
  The AO model fails to accurately predict the entropic attraction for separations $r<2\rsm$ but the MT model remains highly accurate. 
  }
  \label{fig:dep_cWCA}
 \end{center}
\end{figure}

Once again the WCA potential can be subtracted from the total potential $\utot$ in order to produce the depletant-induced pair potential $\udepcWCA$ (\fig{fig:dep_cWCA}). 
The depletion-induced pair potential continues into separations $r<2\rsm$.
As discussed, the definitions of the AO and MT implicitly have infinitely hard cores at $2\rsm$. 
However, this is an imposed restriction and we can remove it without modifying the formalism. 

As seen in \fig{fig:dep_cWCA} (inset), the AO potential crosses the simulated $u$ potential quite near $r\approx2\rsm$ 
but when significant overlap occurs the AO potential $\uao$ cannot reproduce the measured depletion-induced pair potential. 
On the other hand, allowing the MT pair potential $\umtm$ to extend to smaller separations does a remarkable job of reproducing the depletant-induced component of the pair potential except at the highest volume fractions of depletants considered (\fig{fig:tot_cWCA}; inset) at which point the primary repulsive barrier is overpredicted. 
For this reason, summing $\umtm$ and $\ucWCA$ at low to moderate volume fractions once again reproduces the net interaction potentials. 

\begin{figure}[tb]
 \begin{center}
  \includegraphics[width=0.5\textwidth]{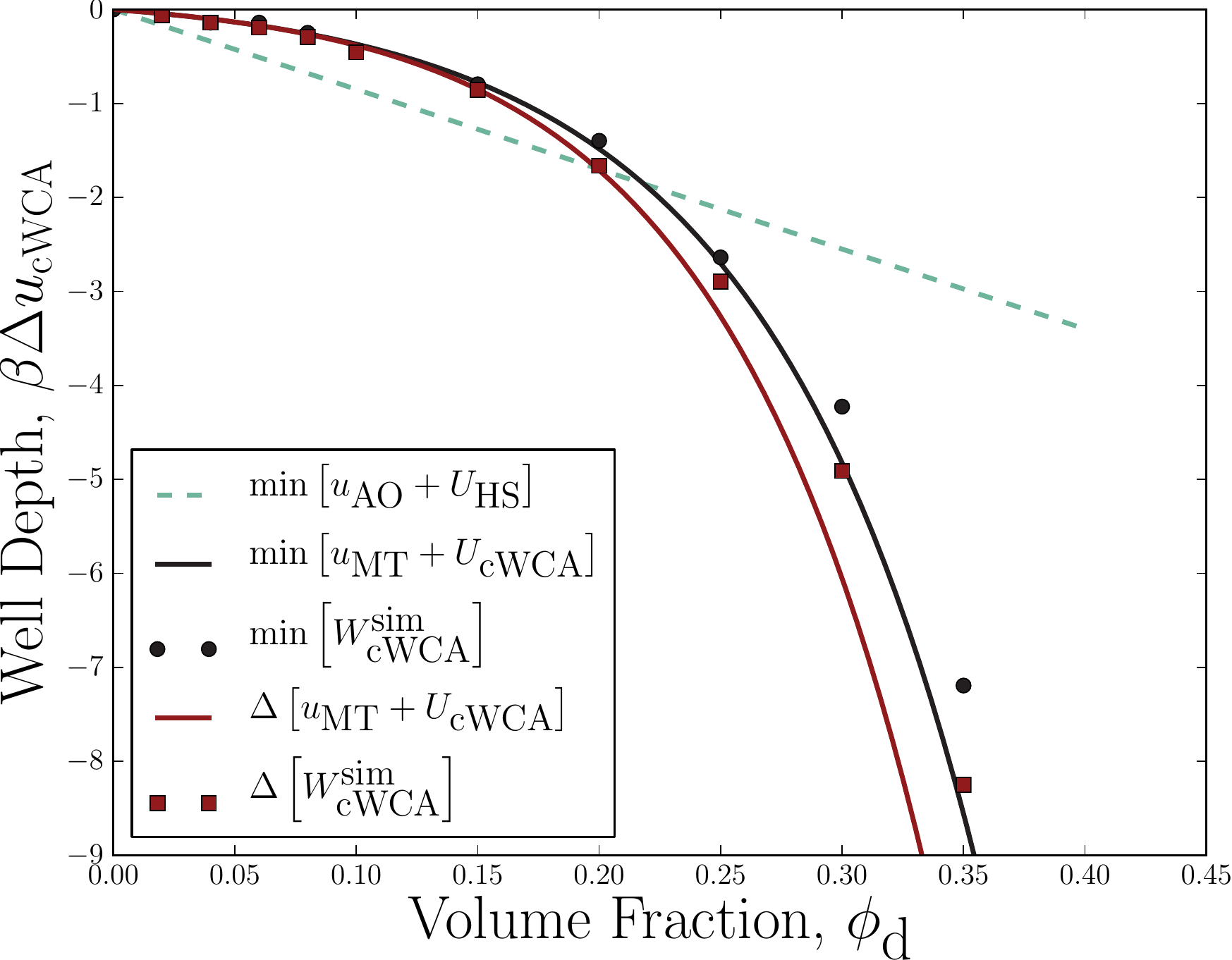}
  \caption{
  The same as \fig{fig:minHard} but for the combinatorial-WCA simulation model. 
  Unlike the shifted-WCA case (\fig{fig:minHard}), the minimum of the Aakura-Oosawa (AO; dashed line) model does not well-approximate the value from simulations at any volume fraction. 
  The theoretical minimum from the MT model plus the combinatorial-WCA potential (solid black line) and the effective well depth are both found to be in good agreement with the
  corresponding simulation result. 
  }
  \label{fig:minSoft}
 \end{center}
\end{figure}

Because the potentials are much softer, greater overlap between the colloids is allowed. 
This in turn yields a significant boost in the volume accessible to the depletants and thus a deep attractive well. 
The resulting minima (\fig{fig:minSoft}) are much deeper at high volume fractions than they were for the sWCA potentials (\fig{fig:minHard}), though not at smaller volume fractions of $\phidep\lesssim0.2$. 
In fact, because the cWCA potential extends further (\fig{fig:pot_plot}), the AO model is not even accurate at low volume fractions. 
At both high and low depletant volume fractions, the AO model for hard-spheres is not sufficient for these cWCA simulations. 
However, since summing the MT model and the cWCA potential reproduces the pair potentials so well for $\phidep\lesssim0.25$ (\fig{fig:tot_cWCA}), it also predicts the behaviour of the minima and the well depths as functions of $\phidep$ quite well (\fig{fig:minSoft}). 
At the highest volume fractions considered ($\phidep\gtrsim0.30$), the MT model once again begins to fail. 
While in the sWCA case, the primary repulsion was overpredicted but the primary well depth was underpredicted (\fig{fig:dep_sWCA} and \fig{fig:minHard}), in the cWCA case \emph{both} the repulsive barrier height and the well depth are overpredicted (\fig{fig:dep_cWCA} and \fig{fig:minSoft}).

\section{Virial Coefficients: Implicit Depletants}
Considering a reduced system of only two colloids in a bath of depletants has allowed us to characterize the depletant pair potential in detail. 
However, systems of interest generally consist of many large particles in a bath of depletants --- a much more computationally expensive system to study. 
To circumvent this, the success of the MT potential in replicating the depletant pair potential indicates that it could be used to include the effects of the depletants \textit{implicitly}. 
In this approach, explicit depletants would not be included in the system but their effects would be included by imposing the MT potential between all pairs of particles.
We use the second virial coefficient as a quantitative measure to predict the efficacy of such an approach.


\subsection{Second Virial Coefficient}\label{2ndvirial}

\begin{figure}[tb]
 \begin{center}
  \includegraphics[width=0.5\textwidth]{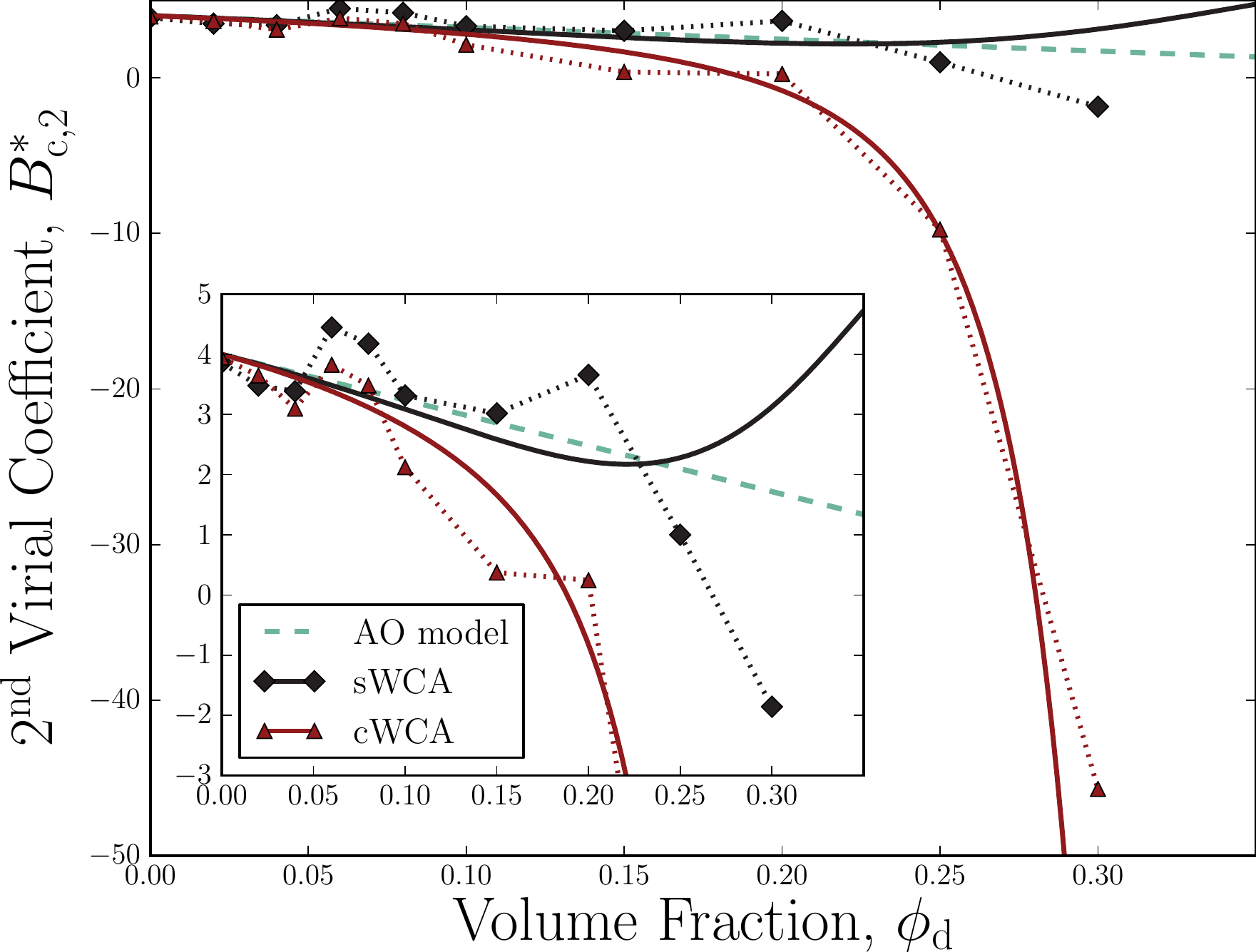}
  \caption{Reduced second virial coefficient $\br=\bsm{2}/\vsm$ for colloidal particles with $\rsm/\rdep=5$ as a function of $\phidep$. 
  Results for the linear Aakura-Oosawa (AO) model is shown as a dashed line. 
  The MT based predictions are also included:  $\umtm + \usWCA$ (solid black line) and   $\umtm + \ucWCA$ (solid red line)  
  }
  \label{fig:B2}
 \end{center}
\end{figure}

Recall that the second virial coefficient was used to calculate the effective size of the colloids according to the potentials that define them.
The effective size as modified by the presence of the depletants can be calculated by a similar method
where the potential $u\left(r\right)$ in \eq{method_2nd} can now be:
i) the simulated pair potentials $\utotsWCA^\textmd{sim}$ and $\utotcWCA^\textmd{sim}$,
ii) the sum of $\umtm$ and $\uWCA$ (either $\usWCA$ or $\ucWCA$),
iii) the net AO potential $\uao+\uhs$. 

Although virial coefficient expansions of the equation of state breakdown in non-thermalized systems, predictions of $\bsm{2}<0$ suggests that attraction dominates over thermal jostling and so demixing of colloidal and depletant particles will occur~\cite{noro00,ashton11}. 
In other words, the colloids are expected to agglomorate for negative enough $\bsm{2}$ values.
The depletion-induced interactions demand a correction such that the virial coefficient is reduced from the hard-sphere value and so we shall discuss the reduced second virial coefficient $\br=\bsm{2}/\vsm$. 

\subsubsection{Simulations}
Consider first the harder, shifted-WCA simulations (\fig{fig:B2}; black diamonds). 
The simulated shifted-WCA second virial coefficient is linear with volume fraction of depletants until approximately $\phidep\approx0.2$ after which point the curvature increases and $\br$ begins to drop more rapidly. 
The second virial coefficient becomes negative somewhere in the range of $\phidep$ between $0.25$ and $0.3$. 

The results for the combinatorial-WCA simulations  (\fig{fig:B2}; red triangles) are similar to the sWCA results at low volume fractions. 
However, the combinatorial-WCA $\br$ falls quite rapidly.
It becomes negative around $\phidep=0.2$ and drops to dramatically more negative values at high volume fraction. 

\subsubsection{AO Approximation: $\phidep \ll 1$}
Assuming that the colloids are hard and $\phidep \ll 1$ simplifies matters immensely. 
This is the AO model and, as shown in \appndx{vanderWaals}, the depletant-induced attractions appear as a correction on the hard-sphere result, specifically taking the form 
\begin{align}
 \frac{\br}{4} &\approx 1 - g \phidep ,
 \label{2ndvirial_AO}
\end{align}
where $g\left(\rdep/\rsm\right)$ is a third order polynomial given by equation \eq{g}. 
The AO approximation (\eq{2ndvirial_AO}) shows that, in this limit, the reduced second virial coefficient decreases as the volume fraction of depletants is increased, reflecting the deepening of the implicit depletion attraction. 
As is expected for the AO model, the correction from the hard-sphere value is linear in $\phidep$. 
The linear AO prediction compares well to both the shifted- and the combinatorial-WCA simulations for $\rsm/\rdep=5$ when the volume fraction of depletants is low (\fig{fig:B2}). 
The AO approximation remains relatively accurate for larger values of $\phidep$ for sWCA. 
For $\phidep \lesssim 0.15$, the approximation predicts the second virial coefficient of the shifted-WCA simulations quite well. 
However, as it must, the AO approximation for $\br$ fails at high volume fractions and does not cross to zero until $\phidep=0.525$ and $0.527$ for sWCA and cWCA, respectively. 

\subsubsection{MT plus WCA Approximation}
While discussing the simulation models, it was shown that summing the MT depletant potential and the WCA excluded volume interactions 
successfully predicts the total pair potentials. 
These theoretical pair potentials can be integrated over to predict the second virial coefficient as a function of volume fraction of depletants.
 
The result for $\umtm + \usWCA$ is shown as a solid black line in \fig{fig:B2}.
The agreement between the AO curve and both simulation results is reasonable until $\phidep\approx0.20$.
At this point, $\br$ for $\umtm + \usWCA$ begins to curve upwards.
This again is an artifact of the exaggeration of the repulsive barrier in the MT model at high volume fractions.
On the other hand, the $\br$ for the cWCA simulations behaves physically, beginning to drop more rapidly at this point. 

At low volume fractions, the result for $\umtm + \ucWCA$ (solid red line) agrees with the data and in fact all other curves.
However, as $\phidep$ increases, the $\br$ for $\umtm + \ucWCA$ begins to drop quite dramatically.
The results are in remarkable agreement with the values derived from the simulations. 
Note that since the cWCA potential allows for the position of the minimum to be significantly less than $2\rsm$ and for correspondingly deeper wells,
the exaggeration of the repulsive barrier at higher volume fractions has a much smaller impact on the calculation of $\br$.

\subsubsection{Escape Time}

The second virial coefficient represents two-body interactions and is often conveniently interpreted as an effective size. 
To give an alternative physical interpretation that is appropriate when the second virial coefficient is a large, negative number, a set of simulations was performed in which two colloids are initially in contact. 
Setting $\rsm/\rdep=5$, the volume fraction of depletants was varied between 0 and 0.35.
Since the depletion force increases with the volume fraction, the particles are held together more tightly with increasing $\phidep$.
To quantify this, the average time it takes for the particles to separate was measured as an escape time, $\tau_\textmd{esc}$.
The colloids are considered to have escaped the depletion potential well when their centre-to-centre distance is larger than $2(\rsm + \rdep)$.

\begin{figure}[tb]
 \begin{center}
  \includegraphics[width=0.5\textwidth]{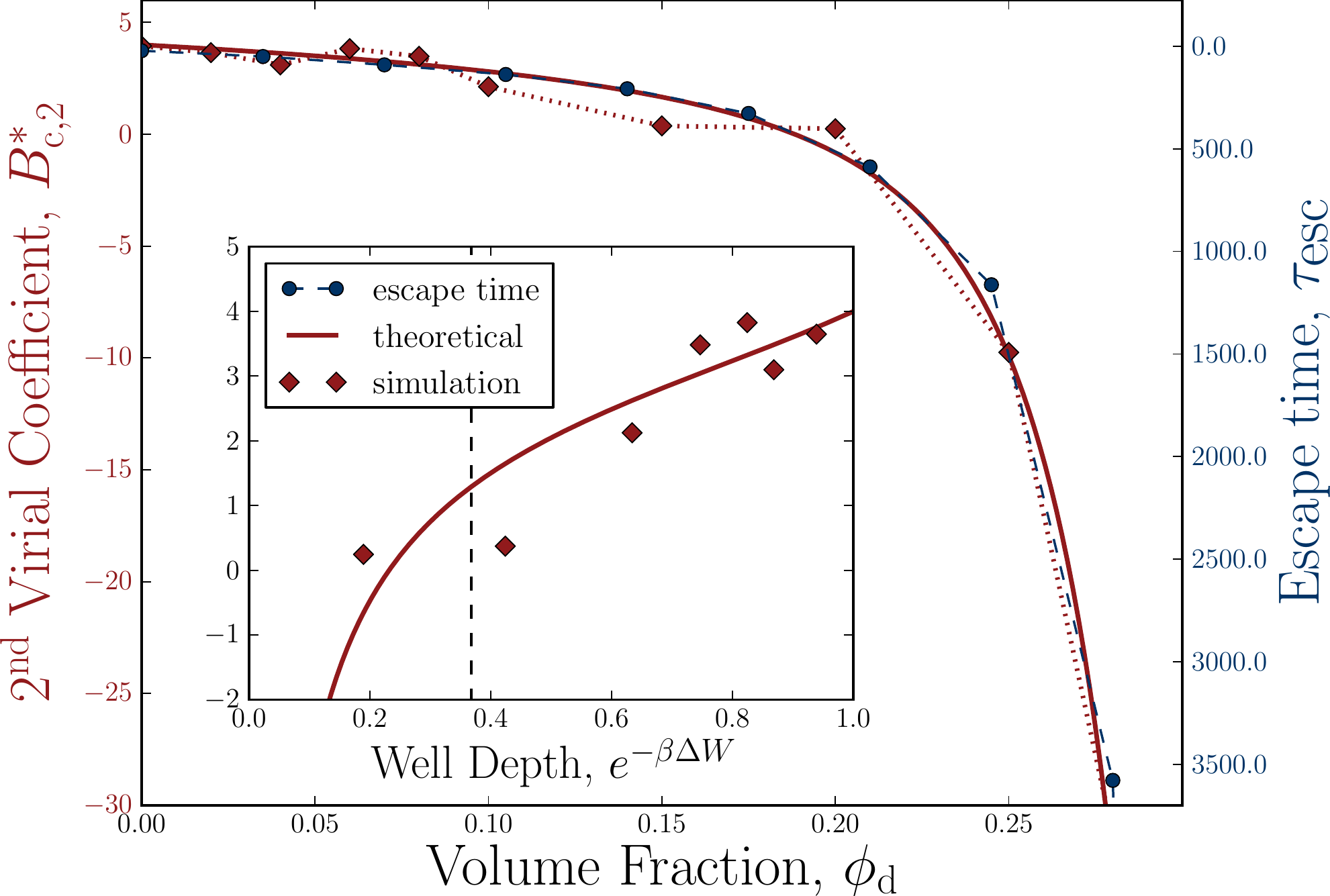}
  \caption{Overlay of the escape time $\tau_\textmd{esc}$ onto the reduced second virial coefficient $\br=\bsm{2}/\vsm$ (triangles denoted cWCA simulations and the solid red line presents the corresponding MT-based predictions) as a function of $\phidep$. 
  The right-hand vertical axis corresponds to the escape time in simulation units. 
  The inset shows the reduced second virial coefficient as an exponential function of the effective well depth, $\Delta W$. 
  }
  \label{fig:B2_tau}
 \end{center}
\end{figure}

As shown in \fig{fig:B2_tau}, the $\tau_\textmd{esc}$ curve mirrors the $\br$ results. 
When the volume fraction of depletants is large the primary attractive well is deep and the primary repulsive barrier is high (\fig{fig:minSoft}) such that colloids that are in contact must diffusively overcome the effective barrier of height $\Delta W$ in order to escape contact. 
In this way, reaching $r=2\mathcal{R}$ is a Kramers' process and $\tau_\textmd{esc} \propto e^{\beta \Delta W}$~\cite{kramers40,hanggi90}. 
As the volume fraction of depletants is increased, the effective barrier height becomes greater (\fig{fig:minSoft}), the escape time rises rapidly as seen in \fig{fig:B2_tau} and the colloids effectively become stuck together. 
Concurrently, the second virial coefficient $\br$ is plummeting to large negative values reflecting the strong attractive interactions as the volume fraction increases. 
The pair potentials (\fig{fig:tot_cWCA}) demonstrate that the primary attractive well and repulsive barrier dominate over the secondary correlations and are indeed the significant contribution to the second virial coefficient. 
The inset to \fig{fig:B2_tau} shows that $\br \sim e^{\beta\Delta W}$  --- as for $\tau_\textmd{esc}$ --- when the effective barrier is greater than the thermal energy $\beta\Delta W>1$. 
The relationship $\tau_\textmd{esc}\sim\br$ suggests a simple physical picture: 
when primary barrier is large the magnitude of $\br$ reflects the relative amount of time that colloids are held together by depletant forces $\tau_\textmd{esc}$. 

\section{Concluding Remarks}

In order to explore coarse-grained methodologies for modelling depletion interactions in soft matter systems,
we charaterized the use of two types of truncated Lennard-Jones (WCA) potentials in Molecular Dynamics simulations of large colloids immersed in a bath of smaller depletants.  

The shifted-WCA model is the ``harder'' of the two models while the combinatorial-WCA potential both extends further and rise less steeply --- and is thus ``softer" (while still being a hard potential).
We find that this seemingly small difference between the excluded volume potentials has a dramatic impact on the depletion effects that arise in the system;
the well depths are deeper, the minimum separations are closer, and the second virial coefficients are much lower for cWCA than sWCA.
Hence, modelling the system with a cWCA approach yields significantly enhanced depletion effects for a fixed volume fraction of depletants. 

Comparison was also made between two simple theories for the depletion potentials.
The simplest approach, the Asakura-Oosawa model for penetrating hard-sphere depletants, was found to give good results in the limit of low volume fractions. 
However, as $\phidep$ increased, this agreement soon broke down for both WCA potentials.
The geometric-based Morphometric Thermodynamics model not only reproduces the minimum in the depletion potential, but also the first repulsive barrier.
Adding $\umtm$ to either $\usWCA$ or $\ucWCA$ yielded excellent agreement for the net pair potentials, the depletion potentials, and the effective well depths at low and moderate volume fractions ($\phidep\lesssim0.25$) but begins to fail for the highest volume fractions considered here. 
However, the second virial coefficient calculated from $\umtm + \ucWCA$ closely matched the cWCA simulation results across all volume fractions, although the predictions for sWCA exhibited known artifacts at these size ratios. 
This result suggests that using the MT depletion potential along side the cWCA excluded volume interaction may be an effective model for \textit{implicitly} incorporating depletion effects into future coarse-grained MD simulations. 

The relative softness of cWCA significantly enhances depletion effects and thus presents a computationally efficient simulation approach since accessible volume fractions generate large primary attractive wells and repulsive barriers. 
Moreover, not only does the use of WCA potentials allow for MD-based exploration of depletant effects but also is likely to yield more physical results for a host of soft matter systems in which interactions are softer than the hard-sphere model. 
In many soft matter systems the effective excluded volume interaction has a softness associated with it; examples include
polymer coated colloids,
solutes that are compressible (e.g. micro or nanogels, dendrimers),
charged objects in an electrolyte solution where the interaction is governed by overlapping Debye layers
or easily deformed materials that can fit together more effectively (e.g., protein folding, chromosome collapse~\cite{part2}).
The results from hard sphere calculations thus represent the lower bound for estimating depletion effects at a given volume fraction and may significantly underestimate their impact for such soft matter systems. 
While the correct modelling will be system specific, this possibility of underestimating depletant effects is important when considering predictions from hard sphere models. 
Our results have shown that even a small amount of softness leading to a marginal amount of overlap can result in a dramatic increase in the depletion potential, resulting in ``soft'' colloids bound together by depletion-induced attraction that would be insufficient for ``harder'' colloids.

\section*{Acknowledgements}
We gratefully acknowledge support through NSERC Discovery Grants to G.W.S. and J.L.H. and EMBO funding to T.N.S (ALTF181-2013).  
High performance computational resources were graciously provided by Prof. Laura Ramuno at the University of Ottawa and Sharcnet. 

\section*{References}
\bibliographystyle{unsrt}
\bibliography{depletantSources}

\appendix

\section{Second Virial Coefficient in the $\phidep\ll1$ Limit:\\Depletant/Colloid van~der~Waals Gas} \label{vanderWaals}
In the $\phidep\ll1$ (or equivalently $\beta\left|\uaomin\right|\ll1$) AO limit, the minimum value of the pair potential is well approximated by the AO model of hard-sphere colloids (\eq{min}) and furthermore the well depth is less than the thermal energy ($\beta\left|\uaomin\right| \ll 1$). 
This limit allows us to approximate the Mayer function $f\left(r\right) = 1-e^{-\beta u\left(r\right)} \approx \beta u\left(r\right)$, which can then be used to find an approximate, analytical form of the second virial coefficient:
\begin{align}
 \bsm{2} 
  &= 2\pi \left[ \frac{2}{\pi} \frac{4 \pi \rsm^3}{3} - \int_{2\rsm}^{2\mathcal{R}}\left(e^{-\beta u}-1\right)r^2 dr + 0\right] \nn\\
  &\approx 2\pi\left[ \frac{2}{\pi} \frac{4 \pi \rsm^3}{3} + \int_{2\rsm}^{2\mathcal{R}} \beta\uao r^2 dr \right] \nn \\
  &= 4\vsm \left[ 1 -  \phidep g \right],
  \label{2ndvirial_AO_appendix}
\end{align}
where we have defined
\begin{align}
 g \left( \frac{\rdep}{\rsm} \right) &\equiv \frac{3}{2}+\frac{15}{8}\left(\frac{\rdep}{\rsm}\right)+\frac{3}{4}\left(\frac{\rdep}{\rsm}\right)^2+\frac{1}{8}\left(\frac{\rdep}{\rsm}\right)^3
 \label{g}
\end{align}
for convenience. 
Notice that setting $\phidep=0$ automatically reproduces the hard-sphere value $\bsm{2} = 4\vsm$. 
The number density of depletants scales the correction, which otherwise is a function of $\rdep/\rsm$ alone (encapsulated in $g$). 
Less intuitively, the effective volume is predicted to reduce when the  size of the depletants relative to the colloids is increased. 
This is a manifestation of the increased range of the attractive well but one must keep in mind the discussion in that the depletants must be significantly smaller than the colloidal particles. 
This condition remains. 

The AO solution can be described as a van~der~Waals gas but, unlike a standard van~der~Waals gas in which only the first term has thermal energy dependence, in an AO solution the pressure scales with $\beta^{-1}$, reflecting the entropic basis of the attractive forces. 

\section{Morphometric Pair Potential Details}

For the MT potential defined in Eqn. \ref{morpho}, the geometric variables are simple functions of colloid-colloid separation $r$~\cite{oettel09}:
\begin{subequations}
  \begin{align}
    V_0 &= \frac{4\pi\mathcal{R}^3}{3}\left[1-\frac{3}{2}\left(\frac{r}{2\mathcal{R}}\right)+\frac{1}{2}\left(\frac{r}{2\mathcal{R}}\right)^3\right], \label{eq:geoa} \\
    A &= 4\pi\mathcal{R}^2\left(1-\frac{r}{2\mathcal{R}}\right), \label{eq:geob} \\
    C_1 &= \frac{A}{\mathcal{R}}+\pi\mathcal{R}\sqrt{1-\left(\frac{r}{2\mathcal{R}}\right)^2}\sin^{-1}\left(\frac{r}{2\mathcal{R}}\right), \label{eq:geoc} \\
    C_2 &= 4\pi. \label{eq:geod}
  \end{align} 
\end{subequations}
However, the thermodynamic quantities are less straightforward. We utilize the Rosenfeld functionals~\cite{rosenfeld89,botan09}:
\begin{subequations}
  \begin{align}
    \beta\Pi\vdep &= -\phidep\frac{1+\phidep+\phidep^2}{\left(1-\phidep\right)^3}, \label{eq:rosena} \\
    \beta\gamma\vdep &= \phidep\rdep\frac{3\phidep\left(1+\phidep\right)}{2\left(1-\phidep\right)^3}, \label{eq:rosenb} \\
    \beta\kappa_1\vdep &= -\phidep\rdep^2\frac{3\phidep^2}{\left(1-\phidep\right)^3}, \label{eq:rosenc} \\
    \beta\kappa_2\vdep &= -\phidep\rdep^3\left(\frac{-2+7\phidep-11\phidep^2}{6\left(1-\phidep\right)^3} - \frac{\ln\left(1-\phidep\right)}{3\phidep}\right), \label{eq:rosend}
  \end{align} 
\end{subequations}
but other accurate forms are available in the binary hard sphere fluid literature such as the White Bear functionals~\cite{roth02,yu02,hansengoos06}. 
The leading term in the osmotic pressure is $\phidep$, while in both the surface tension and $\kappa_2$, the leading terms are $\phidep^2$. 
In $\kappa_1$ the leading term is $\phidep^3$. 

\end{document}